\documentclass [preprint]{aastex}

\shorttitle{Reddest Quasars}

\slugcomment{To appear in the {\em Astrophysical Journal}}

\def\roma#1{\ifmmode{#1}\else{$#1$}\fi} 

\def\kms{\roma{\,\rm km\,s^{-1}\,}}
\def\kmsmpc{\roma{\rm\,km\,s^{-1}\,Mpc^{-1}}}

\newcommand{\MgII}{\ion{Mg}{2}}

\newcommand{\CII}{\ion{C}{2}]}
\newcommand{\CIII}{\ion{C}{3}]}
\newcommand{\CIV}{\ion{C}{4}}
\newcommand{\FeI}{\ion{Fe}{1}}
\newcommand{\FeII}{\ion{Fe}{2}}

\begin{document}

\title {\bf The Reddest Quasars}

\author{
Michael~D.~Gregg\altaffilmark{1,2},
Mark~Lacy\altaffilmark{2,1},
Richard~L.~White\altaffilmark{3},
Eilat Glikman\altaffilmark{4},
David Helfand\altaffilmark{4},
Robert~H.~Becker\altaffilmark{1,2},
Michael~S.~Brotherton\altaffilmark{5}
}

\altaffiltext{1}{Physics Dept., University of California, Davis, CA
95616, gregg,mlacy,bob@igpp.ucllnl.org}
\altaffiltext{2}{Institute for Geophysics and Planetary Physics, L-413
Lawrence Livermore National Laboratory, 7000 East Avenue, Livermore, CA 94550}
\altaffiltext{3}{Space Telescope Science Institute, 3700 San Martin
Drive, Baltimore, MD 21218, rlw@stsci.edu}
\altaffiltext{4}{Columbia Astrophysics Laboratory
eilatg@astro.columbia.edu,djh@carmen.phys.columbia.edu}
\altaffiltext{5}{National Optical
Astronomy Observatory, 950 Cherry Street, Tucson, AZ 85726 mbrother@noao.edu}

\begin{abstract}
In a survey of quasar candidates selected by matching the FIRST and
2MASS catalogs, we have found two extraordinarily red quasars.
FIRST~J013435.7$-$093102 is a 1~Jy source at $z=2.216$ and has $B-K
\gtrsim 10$, while FIRST~J073820.1+275045 is a 2.5~mJy source at
z=1.985 with $B-K \approx 8.4$.  FIRST~J073820.1+275045 has strong
absorption lines of \MgII\ and \CIV\ in the rest frame of the quasar and
is highly polarized in the rest frame ultraviolet, strongly favoring
the interpretation that its red spectral energy distribution is caused
by dust reddening local to the quasar.  FIRST~J073820.1+275045 is thus
one of the few low radio-luminosity, highly dust-reddened quasars
known.  The available observational evidence for
FIRST~J013435.7$-$093102 leads us to conclude that it too is reddened by
dust.  We show that FIRST~J013435.7$-$093102 is gravitationally lensed,
increasing the number of known lensed, extremely dust-reddened quasars
to at least three, including MG0414$-$0534 and PKS1830$-$211.  We discuss
the implications of whether these objects are reddened by dust in the
host or lensing galaxies.  If reddened by their local environment,
then we estimate that between 10 and 20\% of the radio-loud quasar
population is reddened by dust in the host galaxy.  The discovery of
FIRST~J073820.1+275045 and objects now emerging from X-ray surveys
suggests the existence of an analogous radio-quiet red quasar
population.  Such objects will be entirely missed by standard radio or
optical quasar surveys.  If dust in the lensing galaxies is primarily
responsible for the extreme redness of the lensed quasars, then an
untold number of gravitationally lensed quasars are being overlooked.
\end{abstract}

\section {Introduction}

The vast majority of the $>10^4$ catalogued quasars have very similar
optical spectral energy distributions: a very blue continuum with
prominent broad emission lines.  Several studies provide evidence that
a population of {\em red} quasars also exists, overlooked by surveys
optimized to find quasars by their UV excess and strong emission
lines.  The size of any such ``hidden'' population of quasars is
important for a complete understanding of the quasar phenomenon.  Low
et al.\ (1989) found that the majority of quasars in a small
IR-selected sample were significantly redder than optically selected
objects.  Sprayberry \& Foltz (1992) argue that the larger dust
extinction and very weak emission features in low ionization broad
absorption line (BAL) quasars means that they are underrepresented in
the known quasar population.  Using a much larger sample, Webster et
al.\ (1995) have argued that the large range of $B-K$ colors (1 to 8)
of {\it radio-selected} quasars implies that up to 80\% of quasars are
undetected because of several magnitudes of dust obscuration.  Adding
fuel to the fire are the discoveries of striking examples of
individual QSO's with extremely red spectral energy distributions
and/or lacking prominent emission lines.  The continuum of 3CR~68.1
($z = 1.238$) is highly reddened (Brotherton et al.\ 1998) and
polarized.  And the original radio-loud BAL, FIRST~J155633.8+351758
($z = 1.48$; Becker et al.\ 1997), is also quite red with $B-K = 6.3$
(Hall et al.\ 1997) and very weak or absent broad emission lines.
Further evidence of red quasars comes from X-ray surveys.  Dust
reddened quasars are faint or invisible in soft X-rays because of the
high columns of associated absorbing gas (Risaliti et al.\ 2001) but
are relatively bright in hard X-ray surveys now being done with
Chandra (Barger et al.\ 2001).  Countering the dusty quasar
argument are the studies of Benn et al.\ (1998) and Whiting, Webster,
\& Francis (2001) showing that radio-selected quasars are often red
because of an excess of flux at long wavelengths from host galaxy
contributions or a red synchrotron component rather than from
extinction at shorter wavelengths.

Determining the fraction of very red quasars in the total population,
regardless of the cause of their redness, is essential for a precise
determination of the quasar luminosity function, and bears directly on
the evolution of quasars and the nature of the quasar-galaxy
connection.  Also important is whether the red quasar population may
harbor a significant number of optically faint, gravitationally lensed
objects, where extinction in an intervening dusty lens reddens the
quasar light and cancels the magnification bias which accompanies
dust-free lensing (Bartelmann \& Loeb 1999).  A sizable population of
red lensed quasars would have significant ramifications for any
conclusions based on the statistics of gravitational lensing.

The 2-Micron All-Sky Survey (2MASS; Kleinmann et al.\ 1994) provides
an opportunity for gauging the population of very red quasars, and
determining the fraction missed by optically-selected quasar
surveys.  With a resolution of 4\arcsec\ and $10\sigma$ limiting $K_s$
magnitude of 14.3, 2MASS will catalog $>3\times 10^8$ sources with
high-precision coordinates and IR fluxes.  The overwhelming majority
of unresolved 2MASS sources are ordinary Galactic stars, so to use
2MASS effectively in selecting extragalactic point sources requires
additional information.  We have matched 2MASS with the VLA FIRST
Survey 20cm catalog (Becker et al.\ 1995; White et al.\ 1997),
obtaining a sample dominated by extragalactic sources.  Particularly
interesting for the red quasar question is the subset of FIRST/2MASS
matches that are too faint in the optical to appear in extant optical
catalogs such as the APM (McMahon \& Irwin 1992).

We are in the early stages of a program using imaging and spectroscopy
to investigate the nature of these objects.  We report here the
discovery of two extremely red quasars from our sample,
FIRST~J013435.7$-$093102 (also known as PKS~0132$-$097) and
FIRST~J073820.1+275045 (hereafter J0134$-$0931 and J0738+2750,
respectively).  We list their 2MASS magnitudes and FIRST positions and
20cm fluxes in Table~1.  Neither is bright enough optically to appear
in the APM catalog, though both are faintly visible on the digitized
second generation sky survey.  Both appear to be intrinsically red.
One object, J0134-0931, is extremely bright in the IR, which we argue
below is due to gravitational lensing.  The gravitationally lensed
nature of J0134-0931 has been independently discovered by Winn et al.\
(2001a), who investigate its radio properties in detail.  We discuss
whether the redness of these two quasars arises in the host galaxy or
an intervening galaxy along the line of sight, and compare their
properties to other very red quasars in the literature, in particular
MG~0414+0534 (Hewitt et al.\ 1992) and PKS~1830$-$211 (Jauncey et al.\
1991; Courbin et al.\ 1998) which have similarly extreme optical --
near-infrared colors (Table 1).
If even one or two of the lensed objects with extreme spectral energy
distributions is intrinsically red, then the statistics imply that a
significant {\em unlensed} population of red quasars probably exists,
of which J0738+2750 is an example, and the discovery of the two
IR-bright, optically faint quasars in the 2MASS-FIRST sample is
further evidence that optically selected quasars do not provide a
complete sampling of the quasar population.  The intrinsically red
interpretation leads us to speculate that many, perhaps even most,
quasars may undergo a red phase early in their lives when the active
galactic nucleus (AGN) first turns on.  If, on the other hand, the
lensed quasars are reddened by dusty lensing galaxies, then there must
exist numerous undiscovered, optically faint, gravitationally lensed
quasars with small ($< 1\arcsec$) image separations.

\section {Observations}

\subsection{J0738+2750}

J0738+2750 has $K_s = 15.26$ and a FIRST 20cm flux density of 2.6~mJy.
In 1999 November, we obtained an optical spectrum of J0738+2750 using
the Low Resolution Imaging Spectrograph (LRIS, Oke at al.\ 1995) at
Keck Observatory.  The spectrum reveals a very red continuum and
identifiable, but relatively weak, quasar emission features (Figure~1)
at $z = 1.985$.  The \CIV\ emission is particularly weak and both
\MgII\ and \CIV\ show narrow line doublet absorption at or near the
emission line reference frame.  The \MgII\ absorption comprises three
discrete systems at redshifts of 1.97496, 1.98535, and 1.99127, while
the \CIV\ appears to be single and at the emission line redshift.  The
\CIV\ absorption features appear marginally resolved; higher
resolution data may reveal multiple components in the \CIV\
absorption, but at much lower velocity dispersion than in \MgII.  The
velocity dispersion of the three \MgII\ systems is $\sim 500$ \kms\ in
the quasar rest frame.  A broad absorption trough extends from -13000
to -17500 \kms, reaching a depth of $\sim 75\%$ of the continuum
level; we identify this as due to \MgII, making J0738+2750 a weak BAL
quasar.  \FeII\ absorption systems ($\lambda\lambda 2344, 2375, 2383,
2587, 2600$\AA) corresponding to each of the three \MgII\ absorbers
are also present, as is \CII\ emission, which is unusually strong
relative to the other emission lines.  There is also an intervening
\MgII\ + \FeII\ absorption system at $z = 1.4240$.

In 2000 January, we obtained 36 minutes of spectropolarimetry of
J0738+2750, again using LRIS.  These data show that this quasar
highly polarized in the optical, ranging from 5\% at \MgII\ to more
than 9\% at \CIV\ (Figure~1).

In 2000 April, we obtained $J$ and $K'$ band images of J0738+2750
(Figure~2) in photometric conditions using the Near Infrared Imaging
Camera (NIRC, Matthews \& Soifer, 1994) at Keck Observatory.  The 9
point dither pattern totaled 270s of integration time.  Although the
seeing was $0\farcs4$, instrument rotator problems limited the
resolution to roughly twice that.  At this level, J0738+2750 is
unresolved, but 1\farcs8 away is a faint galaxy which is perhaps the
source of the intervening absorption lines.

\subsection{FIRST~J0134-0931}

J0134-0931 is distinguished by being both a strong radio
source (900 mJy at 20 cm in the FIRST survey) and a very bright
infrared object, with 2MASS $K_s = 13.55$.  It is identified in NED as
a quasar, but no redshift is given.  
The original Veron-Cetty \& Veron catalog listed objects as quasars if
they were stellar optical identifications of radio sources, without
requiring spectral confirmation.  The optical ``identification'' of
J0134-0931 is found in Bolton, Shimmens, \& Wall (1975),
the seventh part of the Parkes catalog, where finding charts for
optical counterpart candidates are presented for 122 radio sources.
For 121 of these, the photographic finding charts show an obvious
optical source, but for PKS0132-097, the tick marks enclose blank sky.
Thus, this quasar, ``discovered'' nearly thirty years ago, not only
had no redshift, but also had no optical identification.

In 2000 August, we obtained an infrared spectrum of J0134-0931 at
Lick Observatory using the Gemini IR imaging spectrograph on the Shane
3m telescope.  The spectrum (Figure~3) covers windows in $J, H$, and
$K$, with resolution varying from 30\AA\ in $J$ to 120\AA\ in $K$.  A
single broad emission feature is seen at a wavelength of $2.109\mu$,
which we ascribed to H$\alpha$ at $z = 2.21$.  The observed FWHM of
this line is 280\AA, more than twice the instrumental resolution of
120\AA.  This identification was confirmed by an optical spectrum
obtained with the Andalucia Faint Object Spectrograph Camera (ALFOSC)
on the Nordic Optical Telescope (NOT) at La Palma on 2000 October 30
and 31.  The observations consisted of two 1800s exposures covering
the 4000-9000\AA\ region, and a single 1800s exposure with a different
grism to extend the wavelength coverage redwards to $1\mu$.  Both sets
of observations were made with a 1\farcs2 slit at a PA of 125\arcdeg\
at an airmass of $\sim 1.25$.  The seeing was 0\farcs9 on both nights
and the resolution was 15\AA\ for both spectra.  The optical spectrum
clearly shows emission lines of \CIII\ and \MgII\ at a redshift of $z
= 2.216$ (Figure~3); \CIV\ is weakly detected.

Images of J0134-0931 were obtained with NIRC in $J$ and $K'$ in 2000
September, totaling 270s of integration each.  Conditions were
photometric, the seeing at $2\mu$ was 0\farcs37, and the instrument
rotator problem (which compromised the NIRC observation of J0738+2750)
had been fixed.  The data were reduced using the {\sc dimsum} package
in {\sc iraf} up to the point of sky subtraction, after which the
dithered exposures were subsampled with a spline interpolation by a
factor of 3.14, then shifted and combined.  The $K'$ image (Figure~4)
clearly resolves two components, the brighter component (A) being
extended along one axis in both $J$ and $K'$ bands.  {\sc
daophot/iraf} was used to construct a point spread function (PSF) from
a star near the edge of the image.  Fitting and subtracting the PSF
shows that component D is consistent with a point source while A
leaves large residuals.  (For consistency, we have retroactively
adopted the nomenclature of Winn et al.\ 2001a where the components are
ranked by radio flux; see also below.)  The centroids of the two
objects are 0\farcs635 apart.

A 110s observation of J0134-0931 made on 1992 December 31 with the
Very Large Array (VLA) in the A-configuration at 8.4~GHz (3.6cm) was
recovered from the NRAO archive.  The data were calibrated using
observations of 0141-094 and 3C286 with the {\sc aips} software
package.  The resulting map (Figure~4) has a resolution of 0\farcs3
and is very similar to the NIRC image, showing two components, the
brighter being resolved in one axis at the same position angle as in
the infrared.  Using {\sc aips imfit} to fit elliptical gaussians to
components A and D yields total fluxes of 518 and 69.0~mJy,
respectively, separated by 0\farcs660.  See the accompanying paper by
Winn et al.\ (2001a) for an in-depth discussion of the radio
properties of J0134-0931.


\section {Analysis and Discussion}

\subsection{Lensing of FIRST J0134-0931}

The apparent extremely large luminosity of J0134-0931 initially led us
to suspect that gravitational lensing was at work.  Adopting H$_\circ$ = 70
\kmsmpc\ and $\Omega = 0.3$, the redshift of J0134-0931 implies that
$M_R= -29.6$ even before dereddening.  In the absence of lensing, it
would be one of the most luminous objects in the Universe.  The
morphological similarity of the NIRC K'-band image and the VLA 3.6cm
map virtually confirms that gravitational lensing is at work.

Early in our analysis of J0134-0931, we became aware of the
independent work of Winn et al.\ (2001a) showing that J0134-0931 is a
5-component radio source, which they also interpreted as due to
lensing.  To explore the lensing nature of this object further and to
compare the infrared and radio morphologies, we deconvolved our NIRC
$K'$-band image of J0134-0931 using a variety of techniques.  All
produced qualitatively similar results; here we discuss those
from the maximum entropy task {\sc mem} in {\sc iraf}.  The
deconvolved image is shown in Figure~5 as greyscale, smoothed to
reduce the pixelation, and has an effective resolution of $0\farcs1$
(2 pixels).  The deconvolution breaks component A into two peaks, A
and B (marked by crosses in Figure~5), perhaps with some associated
extended structure at lower surface brightness.  Component D is
consistent with a point source in our $K'$ image.  A fainter and much
less certain source, E, lies to the northeast, plus a hint of excess
flux remaining between components A and D.

We next used the {\sc difmap} package (Shepherd 1997) to map and fit
models to the archival VLA data.  A check of the NASA Extragalactic
Database (NED) shows that J0134-0931 is a GigaHertz-peaked spectrum
(GPS) radio source, and therefore we expect the dominant emission
regions to have a small size compared to the VLA beam (O'Dea,
Baum \& Stanghellini 1991), so we fit models consisting only of point
sources to the data.  Initially we used four point sources: three to
account for the resolved brighter component, plus a fourth for
component D.  A fifth source, C, just to the SE of the brightest
component (A) was required to reduce the otherwise large residuals and
to produce a reasonable fit to the high S/N data.  These five sources
account for 96\% of the flux and there is no evidence for further
discrete components in the residual map, to a limit of $\approx 3$mJy
($\sim 0.4\%$).  The resulting model is shown in Figure~5, convolved
with a $0\farcs1$ restoring beam to match approximately the resolution
of the deconvolved NIRC image; the position of C is indicated by a
black dot.  The four components A, B, C, and E have fluxes of 346, 84,
79, and 20 mJy respectively, while D is 65 mJy.  These ratios are
roughly 100 : 24 : 23 : 19 : 5.8 (A through E).  The radio positions
and flux ratios are consistent with the more detailed, higher
resolution results of Winn et al.\ (2001a).

The radio and IR images in Figure~5 have been placed on the same scale
and registered using component D.  There is good positional agreement
for components A and B; the NIRC data suggest an object consistent
with E, and at still lower significance there is a hint of another
component roughly midway between A and D.  Better signal-to-noise and
resolution are needed to test the reality of these fainter components.

Fitting PSFs at the locations of the five radio components in the
original $K'$ image with {\sc daophot/iraf} greatly reduces the
residuals over the single PSF A-component case, and significantly
improves the fit over those with just two or even three components.
The relative fluxes A:B:C:D:E are, roughly, 100 : 90 : 10 : 25 : 5.
Using the {\sc cplucy} deconvolution task {\sc iraf}, we arrive at
very similar
flux ratios for the 5 component case, 100 : 90 : 14 : 26 : 6.  Given the
limited spatial resolution and depth of the $K'$ data, the results of
fitting five components is evidence only for a qualitative consistency
between the radio and infrared images.

The overall similarity of the $K'$-band and 3.6cm images leads us to
conclude that J0134-0931 is gravitationally lensed, but the exact
nature of the lensing is unclear.  One possibility is a quadruple
image, the 4 components being A, B, D, and E, though the latter is not
unambiguously detected in the NIRC images.  Component C, clearly
detected in the radio and perhaps marginally needed to account for the
2$\mu$ flux, is difficult to interpret, but is perhaps structure in
the quasar radio source and aligned optical emission.
The observed mismatch between the radio and
infrared images can be ascribed to insufficient resolution and
signal-to-noise of the $K$-band data, real optical/radio morphology
differences among the components, or even variability of the radio
source.

One obvious discrepancy between the radio and infrared is the relative
brightness of A and B; though comparable at 2$\mu$, there is a factor
of 4 difference at 3.6cm.  In support of the $K'$ result, similar
analysis with {\sc cplucy} and {\sc daophot} of the somewhat lower
resolution (0\farcs5) NIRC $J$-band image finds that B is at least
80\% as bright as A.  At the same time, the flux ratio of the sum of
the close components (A, B, C, E) to component D is remarkably similar
at 3.6cm (8.1) and $2\mu$ (8.0) especially given the brightness
differences seen among the four closest components.  In the $J$-band
image, we find the A/D brightness ratio to be much greater, $\sim 12$.
At the redshift of J0134-0931, the observed $J$-band is rest frame
near-UV, so the change in color between $J$ and $K'$ of the two
components implies significant differential reddening, or possibly
microlensing effects, or both.

The reader is referred to Winn et al.\ (2001a) for an in-depth
multi-frequency, high resolution radio analysis of the gravitational
lensing of J0134-0931, including a more in-depth discussion of the
lensing model possibilities.

\subsection{Constraining the Reddening}

From the flux calibrated LRIS spectrum of J0738+2750, we obtain $B =
23.8$; combining with the 2MASS data yields a relatively red $B - K
\approx 8.5$.  From our composite optical/IR spectrum of J0134-0931,
we measure $R-K \approx 7.1$ and a {\em lower limit} $B - K \gtrsim
11$ -- there are no detected counts in the $B$ passband.  The extreme
red color of J0134-0931 is confirmed by the broad band photometry of
Winn et al.\ (2001a).
The foreground Galactic extinctions
(Schlegel, Finkbeiner, \& Davis 1998) are small in both cases
(Table~1).  These colors are approximate given the uncertainties of
the spectrophotometric calibration, but the extremely red observed
spectral energy distributions of these objects are beyond question.  A
critical issue is whether the redness of these unusual objects is due
to environmental effects local to the quasar or to intervening
extinction somewhere along our line of sight.

The strong absorption features in the spectrum of J0738+2750 at
the same redshift as the emission lines argue strongly that in this
particular case, the reddening arises locally to the quasar.  The
intervening absorption features are not particularly strong; many
quasars with similar intervening systems are not noticeably reddened
(White et al.\ 2000).  The strong polarization (Figure~1), indicative
of large amounts of local scattering, is also strong evidence that
reddening by dust is responsible for the atypical spectral energy
distribution of this object.  Sprayberry \& Foltz (1992) have found
that BAL quasars are intrinsically redder than non-BAL quasars, so
even the modest \MgII\ BAL (Figure~1) is consistent with local
reddening in J0738+2750.

Determining the origin of the redness of J0134-0931 is more
difficult.  There are no prominent absorption lines in the
optical spectrum and the gravitational lens interpretation raises the
real possibility of intervening reddening and extinction.  Dusty
lenses have been debated as the source of the redness in lensed
quasars (Malhotra, Rhoads, \& Turner 1997; Kochanek et al.\ 2000).

We have attempted to constrain the redshift of the dust using simple
least squares minimization and various popular extinction laws to
redden the composite quasar template spectrum of Brotherton et al.\
(2001) to match the observed spectra of J0738+2750 and
J0134-0931.  The standard Milky Way reddening curve fails because
of the 2175\AA\ silicate ``bump,'' which does not appear in either
source.  The starburst reddening law of Calzetti et al.\ (1994) can
produce a red quasar spectrum that matched the targets, but could not
simultaneously fit the 2MASS photometry points, regardless of the
assumed redshift for the dust.  Only the SMC extinction curve (Pei
1992) could produce results approximating the overall spectral energy
distributions of J0738+2750 and J0134-0931, though still not
with complete success.

Figure~6 compares J0738+2750 with the result of reddening the
composite spectrum by $A_V = 1.78$, assuming the reddening occurs in
the quasar frame.  The 2MASS photometry points were not used to
constrain the fit because of the uncertainty of the scaling between
the slit spectroscopy flux calibration and the IR aperture photometry,
which we estimate to be 10-30\%.  Given the uncertainties of applying
the SMC reddening in this environment, the fit is reasonably good over
the range covered by the Keck spectrum, and fits the 2MASS points
within the overall scaling uncertainty.  There is a continuum slope
mismatch at longer wavelengths, perhaps because of a real difference
between the intrinsic unreddened continuum of J0738+2750 and the
quasar composite.  The observed turndown is a general characteristic
of a reddened power law spectral energy distribution (Francis,
Whiting, \& Webster 2000).  Adjusting the redshift of the extinction
to the intervening absorber system at 1.424 leads to a slight
improvement in the fit over the region covered by the LRIS spectrum,
but the fit is the IR is significantly worse. There is evidence for
excess light in J0738+2750 compared to the reddened quasar spectrum at
short wavelengths ($\gtrsim 4500$\AA), which is probably scattered
quasar light, though a contribution from young stars in the host
galaxy cannot be ruled out.  Based on the high polarization and the
overall consistency of the reddening fits, we conclude that J0738+2750
is reddened by dust local to the quasar.

Figure~7 shows the observed optical-IR spectrum of J0134-0931, with
the 2MASS and $V$ band photometry (Winn et al.\ 2001a) overplotted.
We have scaled the photometry points by a factor of 0.7 by eye to
bring them into agreement with the spectrum; the overall agreement
confirms that the two spectral regions are scaled correctly.  The
$K$-band point drops to the continuum level of the spectrum if
corrected for the presence of the strong H$\alpha$ line.  Two reddened
versions of the composite spectrum are also overplotted.  The solid
line is the composite reddened by $A_V = 2.16$ with dust local to the
quasar.  While the overall shape of the spectrum is reproduced, the
fit does not account for the flux in the $J$-band, and the slope of
the continuum is not reproduced for wavelengths redward of $1.7\mu$.
The dotted line spectrum shows the result of placing the dust at $z =
0.5$, where the best fit using the SMC law yields $A_V = 4.51$.
Moving the dust to lower redshift results in a slightly improved fit
in the IR and perhaps marginally better fit in the optical, but this
is not compelling, given that the same is true for J0738+2750 where
the polarization points conclusively to dust in the quasar
environment.  Lacking additional evidence, such as polarization
measurements or absorption features in the spectrum, we conclude that
our present data are consistent with dust reddening of J0134-0931, but
the redshift of the dust is unconstrained by the modeling.

The existing circumstantial evidence is equally ambiguous as to the
location of the reddening of J0134-0931.
GPS sources like J0134-0931 frequently
occur in systems obscured by large amounts of gas and dust (O'Dea et
al.\ 1996) and where the hosts are often morphologically
``disturbed'', perhaps mergers or interacting galaxies (see review by
O'Dea 1998).  A significant amount of dust associated with the interaction
could 
account for the extreme red energy distribution of J0134-0931.  If
J0134-0931 is a merger or interacting system with large amounts of
dust, then the difficulty in consistently interpreting its IR and
radio morphology (see also Winn et al.\ 2001a) may not be surprising.
The differential reddening that we find in J0134-0931, however, is
highly suggestive of a dusty lens, though microlensing, which may be
needed to explain at least some of the details, cannot be ruled out.
This object is slated for imaging during Cycle~10 of the Hubble Space
Telescope (Falco et al.\, GO9133); the higher resolution optical data
may resolve the situation.  Spectropolarimetry would also help
determine the whereabouts of the dust.  It is, of course, possible
that both the host and the lens are dusty.

The differences between the reddened composite quasar and the observed
red quasars can be ascribed to a multitude of factors.  Strong iron
emission can possibly account for much of the difference in the $J$
band.  The turnover and slope mismatch is perhaps due to intrinsic
differences between the continuum slopes.  There may be host galaxy
contributions with different levels of reddening than the quasars, and
the extinction laws are almost certainly different from the SMC.  An
infrared spectrum of J0738+2750 and a higher S/N infrared spectrum for
J0134-0931 documenting H$\beta$ will permit detailed modeling, as was
done by Lawrence et al.\ (1995a) for the red lensed quasar
MG~0414+0534, and may help clarify these issues.  Determining the
optical colors of the individual components, which will be possible
with the HST images, will also aid in unraveling the nature of
J0134-0931.

To conclude, our data for J0738+2750 place the source of extinction
firmly in the host galaxy.  Our results are somewhat less certain in
the case of J0134-0931.  The lack of strong absorption features in its
spectrum suggests that if the lens is dusty, then it must be at a low
enough redshift to place telltale absorption features far enough into
the blue that they are undetected in our spectrum.  With the presently
available information, we cannot pin down the redshift of the dust for
J0134-0931.

\section{Implications for the Red Quasar Population}

How numerous are dust-reddened objects such as J0134-0931 and
J0738+2750?  Such unusually red quasars suffer so much extinction in
their restframe ultraviolet that they are rendered practically
invisible even at moderate redshifts and are potentially severely
underestimated in optically-selected samples.  Establishing the size
of this red population is of paramount importance to a full
understanding of the general quasar phenomenon and to the
interpretation of quasar gravitational lensing statistics.

Although the very red spectral energy distributions of J0134-0931 and
J0738+2750 are unusual compared to the large majority
of optically-selected quasars, such red objects are much more common
among radio-selected samples of both unlensed (Webster et al.\ 1995;
Francis et al.\ 2000) and lensed quasars (Malhotra et al.\ 1997).
Using matched populations of radio- and optically-selected quasars,
Francis et al.\ (2000) demonstrate that most optically-selected
quasars fall in a narrow color range, $2 > B-K > 3$; roughly 50\% of
radio-selected quasars have similar colors, but the other half are
distributed over the range $3.5 > B-K > 7.5$.  Malhotra et al.\ (1997)
have shown that a similar color dichotomy exists for the known lensed
quasars.  Optically-discovered lensed quasars have broad band colors
consistent with the general quasar population, typically $0 < R-K <
3$, while radio-discovered lensed quasars span a much wider and redder
color range, $\sim 50\%$ having $R-K > 4$, including one of the
reddest extragalactic objects known, MG~0414+0534 with $ R-K \approx
8$ (Lawrence et al.\ 1995a).

A color of $B-K \gtrsim 3.8$ corresponds to an effective
optical-infrared spectral index $\alpha \gtrsim 1$ (defining $\alpha$
in the sense that flux density $S_{\nu} \propto \nu^{-\alpha}$); we
hereafter adopt this our working definition of a ``red quasar''.

\subsection{Red Quasars in Unlensed Samples}

To estimate the fraction of dusty red quasars in the general
population requires an understanding of why the color distribution of
unlensed, radio-selected quasars differs from that of the
optically-selected counterparts.  Benn et al.\ (1998), Francis et al.\
(2000), and Whiting et al.\ (2001) argue that many of the red quasars
in the sample of Webster et al.\ (1995) have intrinsically red spectra
due to optical/near-infrared synchrotron emission from the radio jets
in these relatively bright, flat-spectrum radio sources, but Francis
et al.\ (2000) conclude that dust is necessary to account for a
significant fraction of red quasars, amounting to $\sim 10\%$ of the
total population.  The 3CRR sample of 173 quasars and galaxies (Laing,
Riley \& Longair 1983) contains a number of quasars which have colors
significantly redder than the general population.  As 3CRR is selected
at low radio frequencies, the beamed emission from the radio jet which
can produce strong optical synchrotron emission is much reduced
compared to the sample of Webster et al.\ (1995), and thus these
quasar spectra are unlikely to have significant red optical
synchrotron components.  The fraction of quasars significantly
reddened by dust in 3CRR has been estimated by Simpson et al.\ (1999)
as $\sim 15-50$\%, larger than the fraction deduced by Francis et al.\
(2000), but with small number statistics producing the large error
bars.

Most optical surveys for quasars rely on some form of blue color
selection criterion, so it is not surprising that red quasars
suffering large extinction should be missing from optically-selected
samples.  It is conceivable, however, that the radio-quiet quasar
population simply has, for some reason, no red objects ($B-K > 3.8$).
J0738+2750, although not radio-quiet, is radio-intermediate (if the
optical emission is dereddened), suggesting that this is probably not
the case.  Evidence that such objects may be relatively common comes
from identifications of Chandra hard X-ray sources (Barger et
al.\ 2000) which find significant numbers of lightly-reddened quasars,
and one example of a ``Quasar-2'', in which the broad line region is
completely hidden by dust (Norman et al.\ 2001).  Similarly, using
an objective prism survey and ROSAT data, Risaliti et al.\ (2001) have
identified quasars with anomalously low soft X-ray fluxes, indicating
a significant absorbing column, and find that these quasars are
significantly redder in the optical than the general population.  All
of these results suggest that there is a substantial population of
radio-quieter red quasars, and that the intrinsic color distributions
of radio-loud and radio-quiet quasars may well be similar.

\subsection{Red Lensed Quasars}

The different color distributions of optical- and radio-selected
lensed quasars documented by Malhotra et al.\ (1997) can be explained
largely as a natural consequence of the above discussion -- red
quasars are missing from the optically-selected samples to begin with.
Two other factors enhance the numbers of red, radio-selected lensed
quasars, explaining why Malhotra et al.\ (1997) find that the colors
of radio-loud lensed quasars are even redder than those of the general
radio-loud unlensed population.  First,
a very dusty lensing galaxy can significantly redden the quasar light.
This appears to be the explanation for the very red quasar
PKS1830-211.  The high differential reddening (Courbin et al.\ 1998)
between the components of PKS~1830-211 and the strong differential
molecular absorption at an intervening redshift (Frye, Welch, \&
Broadhurst 1997; Menten et al.\ 1999) is near-conclusive evidence that
the responsible dust is in the lensing galaxy and that the color
differences are not intrinsic to the source or due to microlensing
effects (Courbin et al.\ 1998).

The second factor contributing to the high numbers of red
radio-selected lensed objects is that lensing can magnify the host
galaxy contributions, invariably redder than the quasar light.  Host
galaxy starlight probably explains the very red colors of MG~1131+0456
and B~1938+666 (Kochanek et al.\ 2000).  But Hubble Space Telescope
NICMOS images and near-infrared spectra show that host galaxy
contributions are negligible for the extremely red lensed quasar
MG~0414+0534 (Lawrence et al.\ 1995a), which has colors comparable to
J0134-0931 (see Table 1).  Although Lawrence et al.\ (1995a) conclude
from multicomponent spectral modeling that a dusty lens is responsible
for the red color of MG~0414+0534, the circumstantial evidence argues
that the redness arises from dust local to the quasar.  First, it is a
GPS source.  There are also very strong low-ionization absorption
features of \MgII\ and \FeI\ at the host redshift (Lawrence et al.\
1995b), much like the case of J0738+2750 discussed above, as well as
21cm H{\sc i} absorption with a column consistent with the observed
reddening (Moore, Carilli, \& Menten 1998).  In addition, the lensing
galaxy seems to be a typical elliptical with a low dust content (Tonry
\& Kochanek 1999).

The size of the quasar population which is reddened by dust local to
the source can be estimated from the statistics of radio-discovered
gravitationally lensed quasars.  There are now roughly 18 to 20
gravitationally lensed quasars from various radio searches, including
the Jodrell Bank-VLA Astrometric Survey (JVAS, Patnaik \& Narasimha 1993),
the Cosmic Lens All-Sky Survey (CLASS, Phillips et al.\
2001) and Winn et al.\ (2000; 2001b,c).  Among these are the three
extremely red objects PKS~1830-211 (Courbin et al.\ 1998), MG0414+0534
(Hewitt et al.\ 1992) and J0134-0931 (Winn et al.\ 2001a and this
work).  Assuming that both J0134-0931 and MG0414+0534 are
reddened by dust local to the quasar, then $\sim 10\%$ of the
radio-selected objects are intrinsically red quasars.  

If, however, J0134-0931 and MG0414+0534 are, along with PKS~1830-211,
reddened by their lensing galaxies, the implication is that a
potentially large number of gravitationally lensed quasars exists.  As
only $\sim 15\%$ of quasars are radio-loud, such a population will be
$\sim 7$ times as large as the radio-loud lensed quasars.  As dusty
lenses are likely to be lower mass, late-type galaxies, the image
separations will be small and extinction in the lensing galaxy can
more than counteract any amplification of the quasar light (Bartelmann
\& Loeb 1998).  Such radio-quiet lensed objects are extremely
difficult to discover at any wavelength.

\subsection{Origin of Intrinsically Red Quasars}

Two possible explanations for the existence of intrinsically
dust-reddened quasars are youth and orientation.  Both the GPS and BAL
phenomena have been conjectured to occur early in the lives of quasars
(O'Dea et al.\ 1996; Becker et al.\ 2000).  The reddest quasars,
exemplified by J0134-0931 and MG~0414+0534, both GPS sources, and
J0738+2750, with a BAL feature, may thus be early evolutionary phases
in the lives of quasars, when the AGN is first activated or is
refueled.  Furthermore, Baker \& Hunstead (1997), in a sample of
quasars selected from the Molongolo radio source catalogue found that
the compact steep spectrum (CSS) quasars in their sample typically had
redder than average optical spectra and higher Balmer decrements.
Like GPS sources, CSS sources are thought to be young, with source
evolution proceeding from the GPS through the CSS to the
fully-developed FRII stage (de Vries et al.\ 1998).

Alternatively, within the context of ``unified schemes'' (Antonucci
1993), red quasars may represent objects whose quasar light is just
grazing the dusty torus which hides the quasar nucleus of narrow-line
AGN.  There are well-studied examples from the 3CRR sample which
suggest that youth and orientation are each partly responsible for the
red quasar population: 3C190 is a CSS radio source in a spectacular
merging host galaxy and so possibly a young quasar (Stockton \&
Ridgway 2001), whereas 3C22 and 3C68.1 have all the hallmarks of
evolved radio sources at an angle to the line of sight roughly
corresponding to the hypothesised torus opening angle of $\sim 60$deg.

\section {Conclusions}

If J0134-0931 and MG0414+0534 are intrinsically red, then the
statistics of lensed and unlensed radio-selected
red quasars imply that $\sim
10-20$\% of radio-loud quasars are significantly reddened by dust in
the host galaxy.  The case of J0738+2750 proves that there are at
least some dust reddened quasars with radio luminosities less than the
very high values typical of 3CRR and JVAS/CLASS sources.  
If such intrinsically red objects exist in
similar proportion to the radio-loud red quasars, then a
substantial population of red, radio-quiet objects awaits discovery.
The relative numbers of red quasars and their
redshift distribution can set limits on the relative lengths of
the red phase and its possible relation to the BAL and GPS phases.

If J0134-0931 and MG0414+0534 are reddened by dust in their lensing
galaxies, however, then no large population of intrinsically red
quasars exists.  These objects, along with PKS~1830-211, have been
discovered because they are $\sim 1$~Jy radio sources.  If they are
red because of intervening material along the line of sight, then a
substantial population of radio-quiet, red, gravitationally lensed
systems remains undiscovered, perhaps in numbers 5-10 times as great
as the radio-loud objects.  Their actual numbers are critical to the
interpretation of the statistics of lensed quasars, bearing on
fundamental issues of cosmology and the masses and evolution of
intervening galaxies.

Quasar
surveys utilizing 2MASS and other IR catalogs will eventually build
the statistics of the red quasar population, allowing these issues to
be addressed definitively.
It will be necessary to carry out much of the
exploratory spectroscopy at infrared wavelengths to accurately gauge
the numbers of the most extreme objects.

\acknowledgments

We thank J.~Winn for sharing his radio and optical results for
J0134-0931 ahead of publication and for allowing us to use his broad
band photometry in Figure~7 and Table~1.  He is also thanked for
helpful comments on our manuscript.  Some of the data presented here
were obtained at the W. M. Keck Observatory, which is operated as a
scientific partnership among the California Institute of Technology,
the University of California and the National Aeronautics and Space
Administration.  The Keck Observatory was made possible by the
financial support of the W.M. Keck Foundation.  The FIRST Survey is
supported by grants from the National Science Foundation (grant
AST-98-02791), NATO, the National Geographic Society, Sun
Microsystems, and Columbia University.  Part of the work reported here
was done at the Institute of Geophysics and Planetary Physics, under
the auspices of the U.S. Department of Energy by Lawrence Livermore
National Laboratory under contract No.~W-7405-Eng-48.  The VLA is a
facility of the National Radio Astronomy Observatory, operated by
Associated Universities Inc., under contract with the National Science
Foundation.  The NOT is operated on the island of La Palma jointly by
Denmark, Finland, Iceland, Norway, and Sweden, in the Spanish
Observatorio del Roque de los Muchachos of the Instituto de
Astrofisica de Canarias. ALFOSC is owned by the Instituto de
Astrofisica de Andalucia (IAA) and operated at the NOT under agreement
between IAA and the NBIfAFG of the Astronomical Observatory of
Copenhagen.  K.~Brand assisted with the NOT observations.  
The Two Micron All Sky Survey (2MASS) is a
joint project of the University of Massachusetts and the Infrared
Processing and Analysis Center/California Institute of Technology,
funded by the National Aeronautics and Space Administration and the
National Science Foundation (NSF).  This research has made use of the
NASA/IPAC Extragalactic Database (NED) which is operated by the Jet
Propulsion Laboratory, California Institute of Technology, under
contract with the National Aeronautics and Space Administration.


\pagebreak

\begin {deluxetable}{lllcllrrrrrrl}
\tabletypesize{\scriptsize}
\tablewidth{0in}
\tablehead{
\colhead{Name} &
\colhead{$\alpha\tablenotemark{a}$ (J2000) } &
\colhead{$\delta\tablenotemark{a}$ } &
\colhead{Gal.\ $A_V$} &
\colhead{$z$} &
\colhead{$B$} &
\colhead{$V$} &
\colhead{$R$} &
\colhead{$I$} &
\colhead{$J$} &
\colhead{$H$} &
\colhead{$K$} &
\colhead{$S_{20cm}$ (mJy)}
}
\startdata
FIRST~0134$-$0931&  
$\mathrm{ 01^{h} 34^{m} 35\fs7 }$  &
 $-09\arcdeg 31\arcmin 03\arcsec$  &
 0.10 &
 2.216 &
 $>24.3$  & 
 22.57   & 
 20.61   & 
 18.77   & 
 16.17  & 
 14.75  & 
 13.55   & 
 900.  \\
FIRST~0738$+$2750 &  
$\mathrm{ 07^{h} 38^{m} 20\fs1 }$  &
 +27\arcdeg 50\arcmin 46           &
 0.14 &
 1.985 &
 23.8  & 
 22.1   & 
 20.6   & 
 19.2   & 
 17.1   & 
 16.2   & 
 15.3   & 
 2.6  \\
MG~0414$+$0534&  
$\mathrm{ 04^{h} 14^{m} 37\fs8 }$  &
 $+05\arcdeg 34\arcmin 42\arcsec$  &
 0.10 &
 2.639 &
 24.1  & 
 23.8   & 
 21.8   & 
 19.5   & 
 15.7   & 
 14.2  & 
 13.7   & 
 $2676$  \\
PKS~1830$-$2111&  
$\mathrm{ 18^{h} 33^{m} 39\fs9 }$  &
 $-21\arcdeg 03\arcmin 40\arcsec$  &
 1.54 &
 2.507 &
 ---  & 
 ---   & 
 ---   & 
 21.4   & 
 18.7   & 
 ---  & 
 15.0   & 
 $\sim 10^{4}$  \\
\enddata
\tablenotetext{a} {FIRST survey (Becker et al.\ 1995)}
%
\tablecomments{Optical photometry for J0134-0931 from Winn et al.\
2001, for J0738+2750 from our flux calibrated spectrum, for
MG~0414+0534 from Lawrence et al.\ (1995a).  \\
IR photometry from 2MASS, Lawrence et al.\ (1995a), or Courbin et al.\
1998.  \\
Radio flux density from FIRST survey or NED.}

\end {deluxetable}

\clearpage

\begin{figure}[t]
\plotone{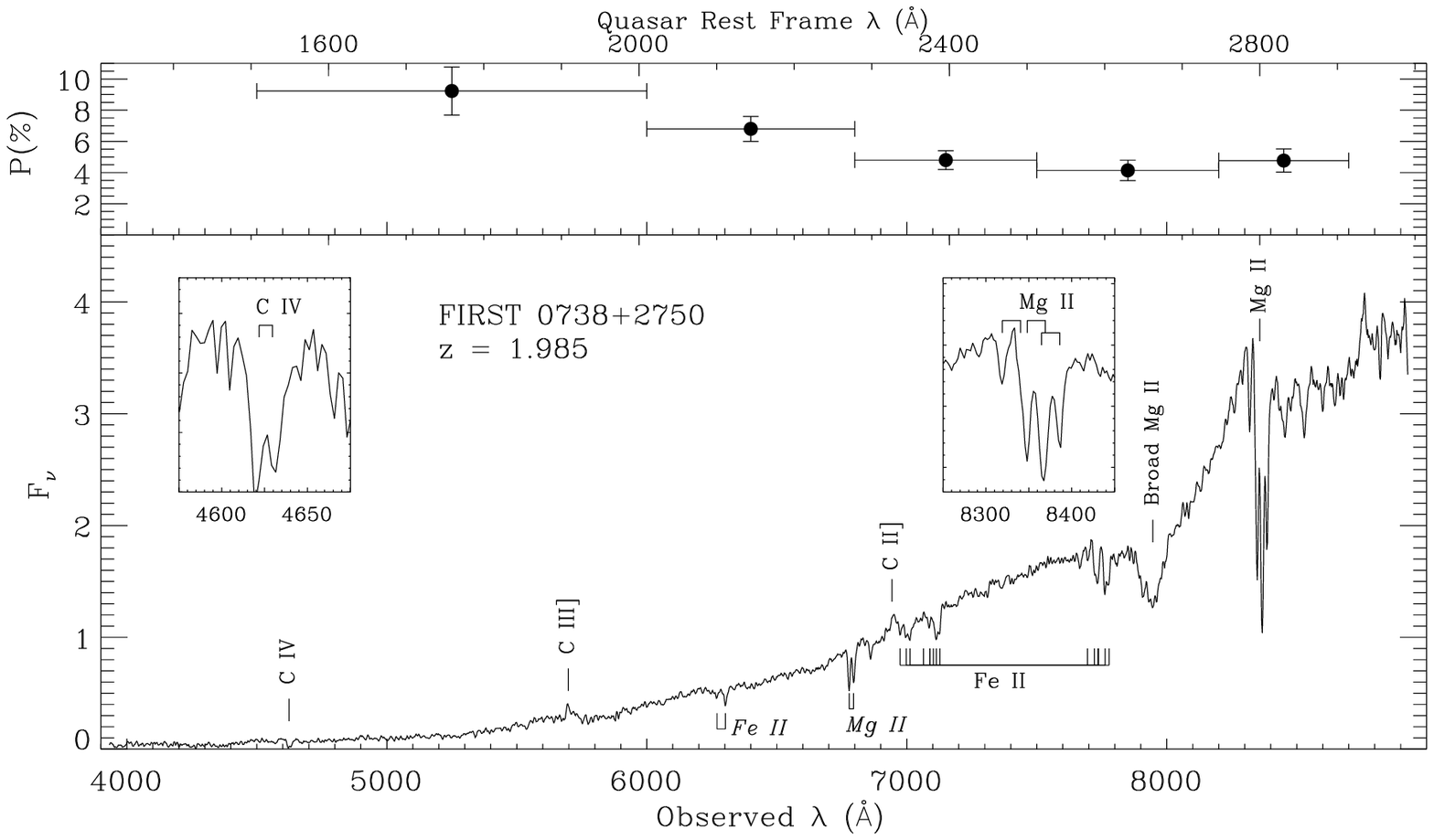}
\end{figure}
\noindent
{Fig.\ 1.-- Flux-calibrated Keck LRIS spectrum of J0738+2750 showing the
very red continuum.  The \MgII\ 2800 and \CIV\ 1549 regions have
narrow line absorption at the emission line redshift and the \MgII\
absorption is triple with a velocity dispersion of $\sim 500$ \kms.
\FeII\ absorption corresponding to each of these is also present, as
is a broad \MgII\ absorption feature extending from $-13000$ to
$-17500$ \kms\ in the quasar restframe.  \newline There is also an
intervening absorber contributing narrow \MgII\ and \FeII\ lines at $z
= 1.424$ (in italics).  The polarization data, also from Keck LRIS,
show an increase from 4\% to more than 9\% between \MgII\ and \CIV,
supporting the interpretation that the unusual spectral energy
distribution is due to reddening by dust.  The polarization error bars
show $2\sigma$ errors; wavelength error bars indicate the binning used
to measure the polarization signal.}

\clearpage
\setcounter{figure}{1}
\begin{figure}
\epsscale{0.5}
\plotone{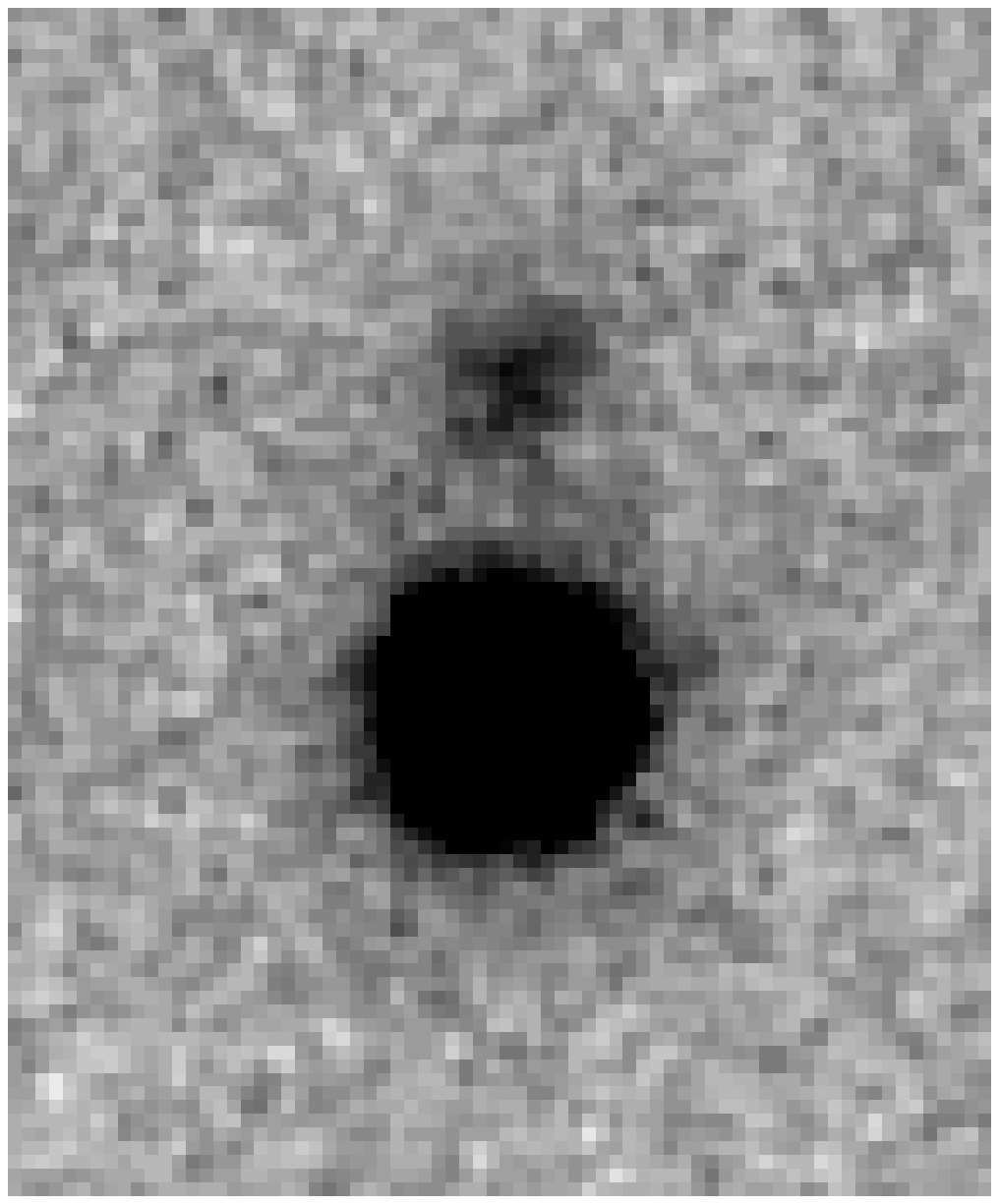}
\caption{Keck NIRC $K'$ image of J0738+2750; north is up, east to
the left.  The image quality
is limited to $\sim 0\farcs8$ by intrument rotator errors; the source
appears unresolved at this level.  The faint galaxy 1\farcs8 North
may be the source of the intervening absorption lines at $z = 1.424$.
}
\end{figure}

\clearpage

\begin{figure}
\epsscale{1.}
\plotone{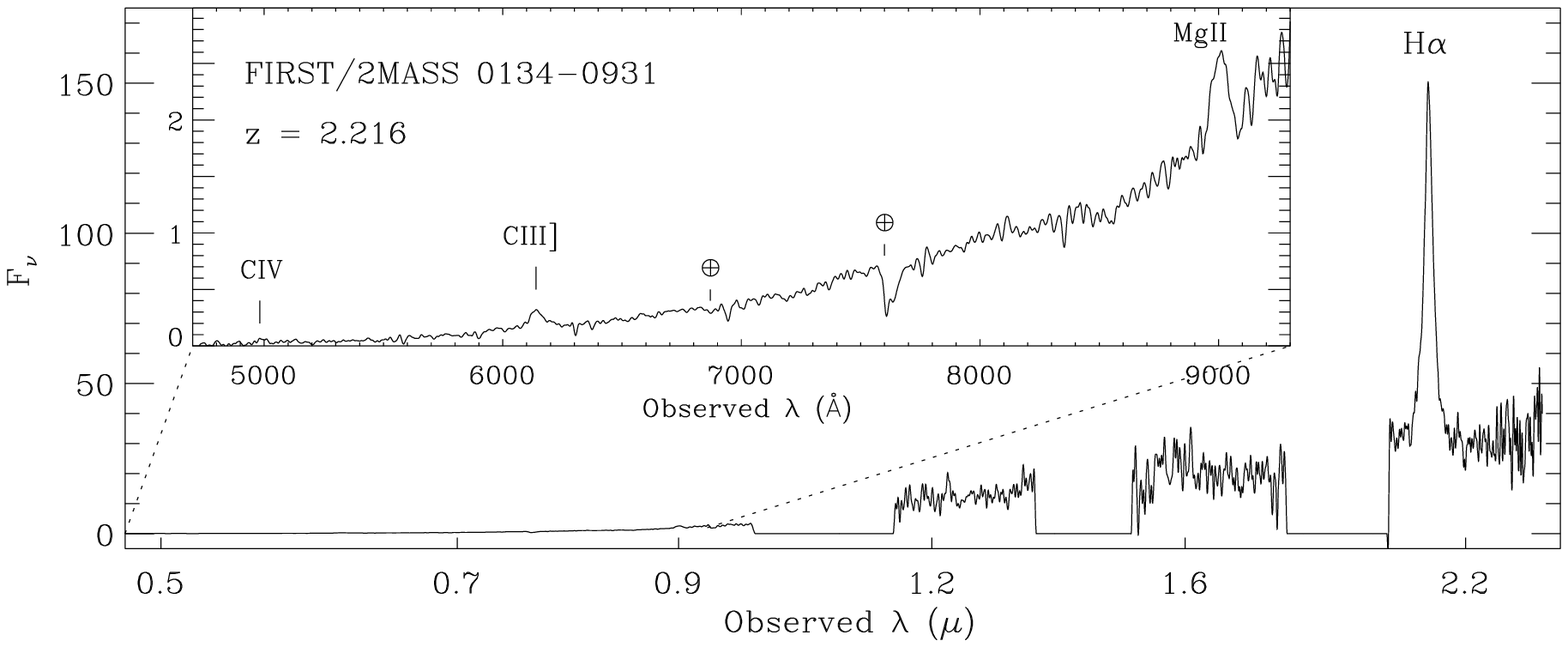}
\caption{Combined Lick 3m IR and NOT optical spectra of
J0134-0931 showing the remarkably red overall spectral energy
distribution.  }
\end{figure}

\clearpage

\begin{figure}
\plotone{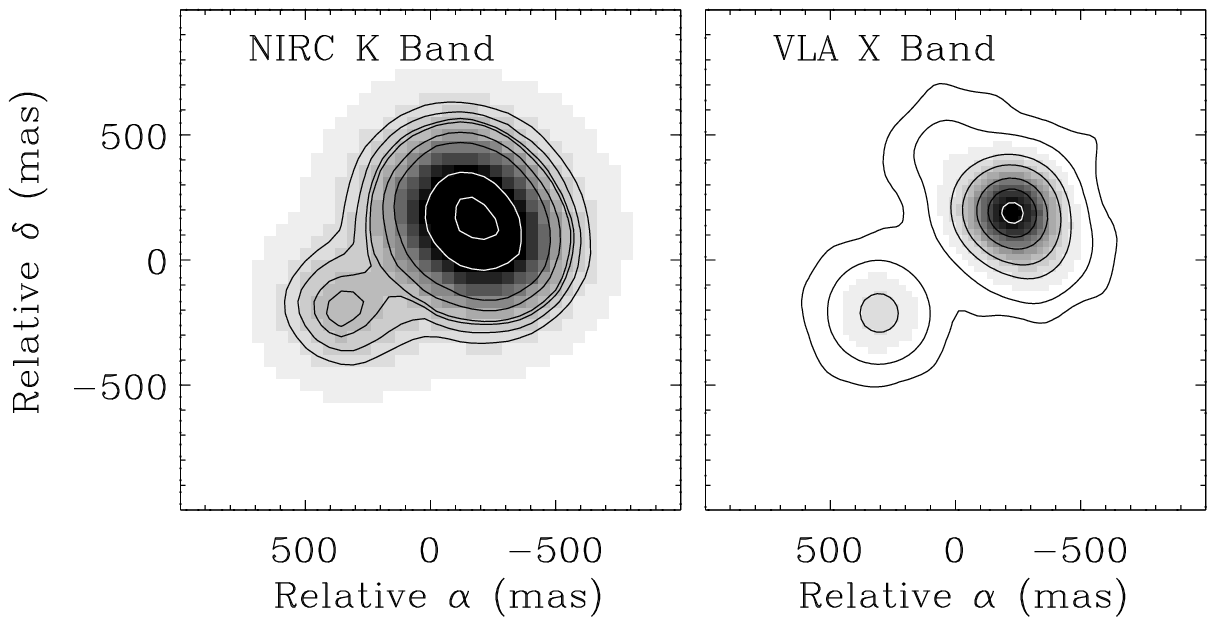}
\caption{Comparison of Keck NIRC $K'$ image obtained in $0\farcs37$
seeing and archival VLA A-array 3.6cm map of J0134-0931.  The
overall similarity of the two strongly suggests gravitational lensing.
}
\end{figure}

\clearpage

\begin{figure}
\plotone{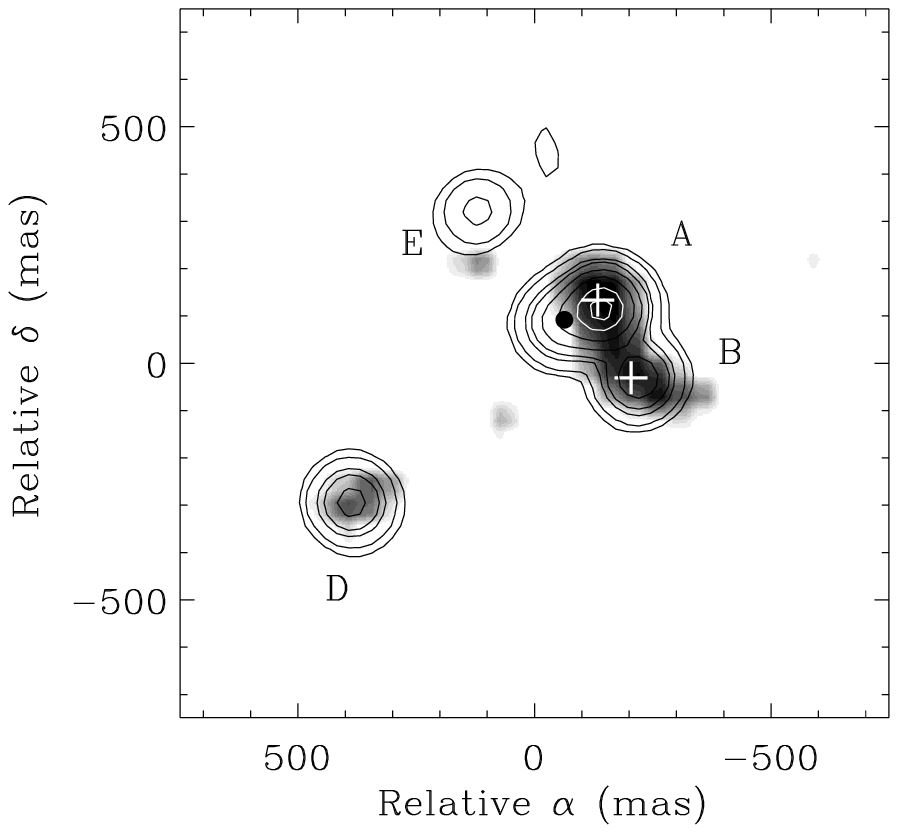}
\caption{Keck NIRC $K'$ of J0134-0931 after maximum entropy
deconvolution (greyscale) compared with the enhanced resolution {\sc
difmap} image (contours) obtained by modeling the archival VLA A-array
3.6cm data with point sources; the position of the extra component,
C, is indicated by the black dot.  The two datasets have been aligned
using component D.  The white crosses mark the locations of the two
brightest peaks in the deconvolved $K'$ image, showing good positional
agreement with the radio peaks.}
\end{figure}

\clearpage

\begin{figure}
\plotone{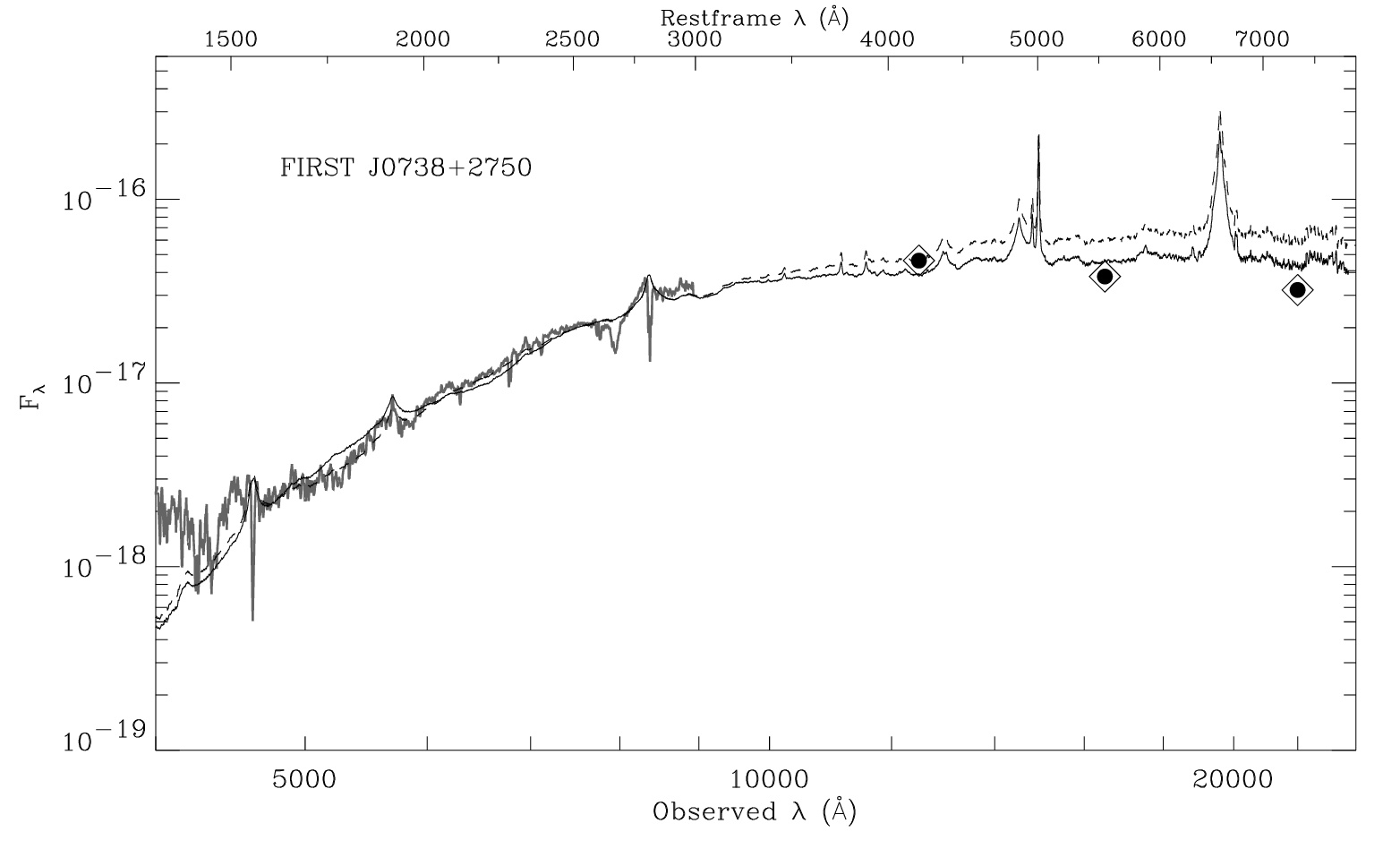}
\caption{Comparisons of J0738+2750 Keck spectrum (thick grey
line) and 2MASS JHK photometry with a
reddened version of the quasar composite spectrum from Brotherton et
al.\ (2001).  The SMC extinction curve (Pei 1992) was used with
$A_V=1.78$. 
The excess observed flux at short
wavelengths is consistent with the amount of scattering needed to
produce the polarization (see Figure~1). }
\end{figure}

\clearpage

\begin{figure}
\plotone{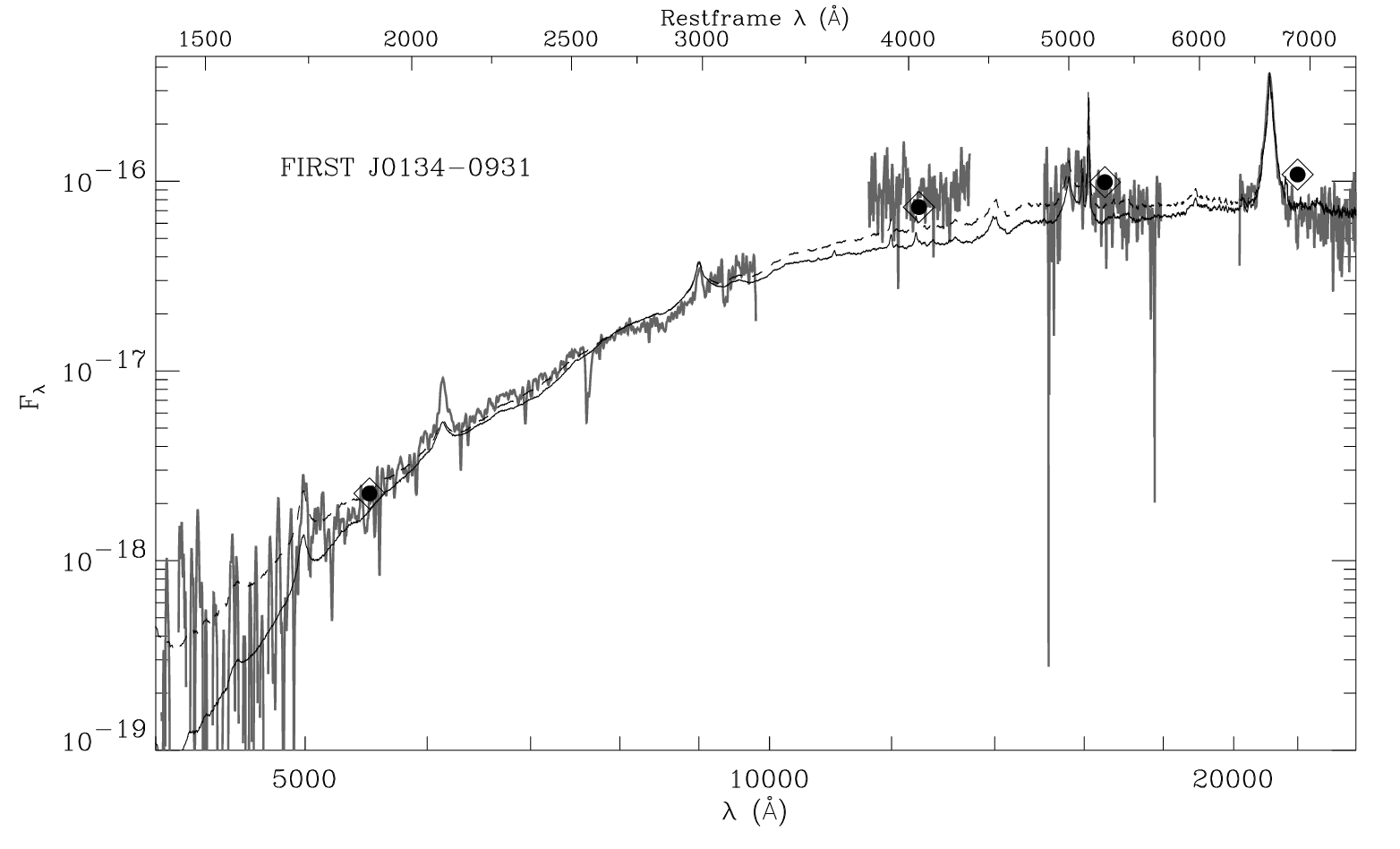}
\caption{Comparison of J0134-0931 with reddened versions of the
quasar composite spectrum from Brotherton et al.\ (2001).  The thick
grey line is the observed spectrum of J0134-0931; photometry
points from 2MASS and Winn et al.\ (2001) are overplotted.  The thin
black line is the composite reddened by $A_V = 2.16$ using the SMC
extinction curve (Pei 1992) with the dust in the rest frame of the
quasar.  Dashed line is for $A_V = 4.51$ and the dust at $z = 0.5$.
}
\end{figure}


\begin{references}

\reference{} Annis, J. \& Luppino, G., A. 1993, ApJL, 407, L69

\reference{} Antonucci R., 1993, ARA\&A, 31, 473

\reference{} Baker J.C., Hunstead R.W., 1995, ApJ, 452, L95

\reference{} Barger, A. J., Cowie, L. L., Mushotzky, R. F., \&
Richards, E. A. 2001, AJ, 121, 662

\reference{} Bartelmann, M. \& Loeb, A.  1998, ApJ, 503, 48

\reference{} Becker, R. H., White, R. L., \& Helfand, D. J. 1995, ApJ,
450, 559

\reference{} Becker, R. H., White, R. L., Gregg, M. D., Brotherton,
M. S., Laurent-Muehleisen, S. A., \& Arav, N. 2000, ApJ, 538, 72

\reference{} Becker, R. H., Gregg, M. D., Hook, I. M., McMahon, R. G.,
White, R. L., \& Helfand, D. J.  1997, ApJL, 479, L93

\reference{} Benn, C. R., et al.\ 1998, MNRAS, 295, 451

\reference{} Bolton, J. G., Shimmins, A. J., \& Wall, J. V. 1975,
Australian J. Phys. Ap. Suppl., Vol.\ 34, 1

\reference{} Brotherton, M. S., van Breugel, W., Smith, R. J., Boyle,
B. J., Shanks, T., Croom, S. M., Miller, L. \& Becker, R. H.  1998, ApJL,
505, L7

\reference{} Brotherton M.S., Wills B.J., Dey A., van Breugel W., 
Antonucci R., 1998, ApJ, 501, 110

\reference{} Brotherton, M. S., Tran, H. D., Becker, R. H., Gregg,
M. D., Laurent-Muehleisen, S. A., \& White, R. L.  2001, ApJ, 546, 775

\reference{} Calzetti, D., Kinney, A. L., Storchi-Bergmann, T. 1994,
ApJ, 429, 582

\reference{} Courbin, F., Lidman, C., Frye, B. L., Magain, P.,
Broadhurst, T. J., Pahre, M. A., \& Djorgovski, S. G. 1998, ApJL, 499,
L119

\reference{} de Vries W.H., O'Dea C.P., Perlman E., Baum S.A., Lehnert M.D., 
Stocke J., Rector T., Elston R., 1998, ApJ, 503, 138 

\reference{} Economou F., Lawrence A., Ward M.J., Blanco P.R., 1995, MNRAS, 
272, L5

\reference{} Falco, E. E., et al.\ 1999, ApJ, 523, 617

\reference{} Francis, P., Whiting, M., \& Webster, R. 2000, PASA, 53,
56

\reference{} Frye, B., Welch, W. J., \& Broadhurst, T. 1997, ApJL,
478, L25

\reference{} Hall, P. B., Martini, P., DePoy, D. L., \& Gatley, I.
1997, ApJL, 484, L17

\reference{} Hewitt, J. N., Turner, E. L., Lawrence, C. R., Schneider,
D. P., Brody, J. P. 1992, AJ, 104, 968

\reference{} Jauncey, D. L. et al.\ 1991, Nature, 352, 132

\reference{} Kleinmann, S.G., Lysaght, M.G., Pughe, W.L., Schneider,
S.E., Skrutskie, M.F., Weinberg, M.D., Price, S.D., Matthews, K.,
Soifer, B.T., \& Huchra, J.P. 1994, ApJS, 217, 11 

\reference{} Kochanek, C. S., et al.\ 2000, ApJ, 535, 692

\reference{} Laing R.A., Riley J.M., Longair M.S., 1983, 204, 151

\reference{} Lawrence, C. R., Elston, R., Januzzi, B. T., \& Turner,
E. L. 1995a, AJ, 110, 2570

\reference{} Lawrence, C.R., Cohen, J. G., \& Oke, J. B. 1995b, AJ,
110, 2583

\reference{} Low, F.J., Cutri, R.M., Kleinman, S.G., \& Huchra,
J.P. 1989, ApJL, 340, 1

\reference{} Malhotra, S., Rhoads, J.E., \& Turner, E.L. 1997, MNRAS,
288, 138

\reference{} Matthews, K., \& Soifer, B. T. 1994, in Infrared
Astronomy with Arrays: the Next Generation, ed.  I. McLean (Dordrecht:
Kluwer), 239

\reference{apm} McMahon, R.G., \& Irwin, M.J. 1992, in {\it Digitized
Optical Sky Surveys}, eds.~H.T. MacGillivray and E.B. Thomson, Kluwer,
p.~417

\reference{} Menton, 1999

\reference{} Moore C.B., Carilli C.L., Menten K.M., 1998, ApJ, 510, L87

\reference{} Norman C., et al.\ 2001, ApJ, submitted (astro-ph/0103198)

\reference{} O'Dea, C. P., Baum, S. A., \& Stanghellini, C. 1991, ApJ, 380, 660

\reference{} O'Dea, C. P., Worrall, D. M., Baum, S. A., \&
Stanghellini, C. 1996, AJ, 111, 92

\reference{} O'Dea, C. P. 1998, PASP, 110, 493

\reference{} Oke, J. B., Cohen, J. G., Carr, M., Cromer, J., Dingizan,
A., Harris, F. H., Labrecque S., Lucinio, R., Schaal, W., Epps, H., \&
Miller, J. 1995, PASP, 107, 375

\reference{} Patnaik, A. \& Narasimha, D.  1992,
Bull. Astr. Soc. India, 21, 457

\reference{} Pei, Y. C. 1992, ApJ, 395, 130

\reference{} Phillips, P. M., et al.\ 2000, astro-ph/0009334


\reference{} Rawlings S., Lacy M., Eales S. A., Sivia D. S., 1995, MNRAS, 274,
428

\reference{} Risaliti G., Marconi A., Maiolino R., Salvati M., Severgnini
P., 2001, A\&A, in press 

\reference{} Schlegel, D.J, Finkbeiner, D.P., \& Davis, M. 1998, ApJ,
500, 525

\reference{} Shepherd, M.C.  1997, in ASP Conf.~Ser.~125, Astronomical
Data Analysis Software and Systems VI, eds. G. Hunt, H. E. Payne, 77).

\reference{} Simpson C., Rawlings S., Lacy M., 1999, MNRAS, 306, 828

\reference{} Smith H.E., Spinrad H., 1980, ApJ, 236, 419

\reference{} Sprayberry, D., \& Foltz, C. B. 1992, ApJ, 390, 39

\reference{} Stockton A., Ridgway S. E., 2001, ApJ, in press (astro-ph/0102293)

\reference{} Tonry J. L. \& Kochanek C. S., 1999, AJ, 117, 203

\reference{} Veron-Cetty, M. P. \& Veron, P. 1983, A\&AS, 53, 219

\reference{} Webster, R.L., Francis, P.J., Peterson, B.A., Drinkwater, M.J.,
\& Masci, F.J. 1995, Nature, 375, 469

\reference{} White, R. L., Becker, R. H., Helfand, D. J., \& Gregg,
M. D. 1997, ApJ, 475, 479

\reference{} White, R.L., et al.\ 2000, ApJS, 126, 133

\reference{} Whiting, M., Webster, R., \& Francis, P. 2001,
astro-ph/0101502

\reference{} Willott C.J., Rawlings S., Blundell K.M., Lacy M., 2000, 
MNRAS, 316, 449

\reference{} Winn, J. et al.\ 2000b, AJ, 120, 2868

\reference{} Winn, J. et al.\ 2001a, ApJ, 000, 000

\reference{} Winn, J. et al.\ 2001b, astro-ph/0104092

\reference{} Winn, J. et al.\ 2001c, AJ, 121, 1223

\end{references}
\end{document}